\documentclass{article}

\usepackage{arxiv}

\usepackage[utf8]{inputenc} 
\usepackage[T1]{fontenc}    
\usepackage{hyperref}       
\usepackage{url}            
\usepackage{booktabs}       
\usepackage{amsfonts}       
\usepackage{nicefrac}       
\usepackage{microtype}      
\usepackage{lipsum}		
\usepackage{graphicx}
\usepackage{doi}

\usepackage{amssymb,amsmath,url}

\usepackage{multirow,caption}

\usepackage[authoryear]{natbib}

\newtheorem{assumption}{Assumption}

\title{On the Use of Two-Way Fixed Effects Models for Policy Evaluation During Pandemics}

\date{\today} 					

\author{Germain Gauthier\\
	Ecole Polytechnique (CREST) \\
	Route de Saclay, Palaiseau, France \\
    \texttt{germain.gauthier@polytechnique.edu}
}

\renewcommand{\shorttitle}{Two-Way Fixed Effects Models During Pandemics}

\hypersetup{
pdftitle={Two-Way Fixed Effects Models During Pandemics},
pdfsubject={Econ},
pdfauthor={Germain Gauthier},
pdfkeywords={Covid-19, difference-in-differences, policy evaluation},
}

\begin{document}
\maketitle

\begin{abstract}
        In the context of the Covid-19 pandemic, multiple studies rely on two-way fixed effects (FE) models to assess the impact of mitigation policies on health outcomes. Building on the SIRD model of disease transmission, I show that FE models tend to be misspecified for three reasons. First, despite misleading common trends in the pre-treatment period, the parallel trends assumption generally does not hold. Second, heterogeneity in infection rates and infected populations across regions cannot be accounted for by region-specific fixed effects, nor by conditioning on observable time-varying confounders. Third, epidemiological theory predicts heterogeneous treatment effects across regions and over time. Via simulations, I find that the bias resulting from model misspecification can be substantial, in magnitude and sometimes in sign. Overall, my results caution against the use of FE models for mitigation policy evaluation.
\end{abstract}


\section{Introduction}

The current Covid-19 pandemic poses a major and global health issue. As most political interventions to contain the pandemic involve an economic slowdown and a restriction of individual liberties, governments are facing an uneasy trade-off between the health benefits and the costs that these policies entail. Lockdown policies, in particular, come at a high economic and social cost. Given the stakes, rigorously assessing the effectiveness of such policies is critical. 

An increasingly large body of evidence suggests that mitigation policies have slowed down the growth of Covid-19 infections. However, it remains unclear how much of the observed slow down in the spread is attributable to the effect of policies as opposed to other confounding factors. For instance, individuals could apply to themselves self-distancing measures out of fear of being infected. They could also learn more about the disease, its seriousness and the appropriate means of protection as the epidemic unfolds. 

To alleviate these concerns, multiple studies in the economics literature exploit the staggered implementation of mitigation policies across regions to identify treatment effects. This is a familiar setup for empirical work in economics, which intuitively suggests the use of difference-in-differences (DID) methodology. In practice, researchers rely on two-way fixed effects (FE) linear regressions. They thus model a transform $g(\cdot)$ of confirmed infected cases (e.g. $\text{log(confirmed cases)}$) as a linear sum of time fixed effects, region fixed effects, a mitigation policy implementation and potentially several control variables.

In this paper, I study whether FE models are consistent with the Susceptible-Infected-Recovered-Deceased (SIRD) epidemiological model of disease transmission. My main result is that they tend to be misspecified for three reasons: (a) the violation of the parallel trends assumption, (b) the inadequacy of the assumed data-generating process, and (c) the presence of heterogeneous treatment effects (over time and across regions). As a direct consequence, FE models generally result in biased treatment effect estimates. I discuss each of these issues in more detail below.

I start with the SIRD model and derive a simple framework to think of confounding factors in the context of an epidemic.  The observable dependent variable -- i.e., confirmed cases (or alternatively death counts) -- is driven by three unobservables: the size of the infected population, the size of the susceptible population and the infection rate between individuals. From this observation, it becomes clear that there are two main types of confounding factors which complicate causal inference: (a) heterogeneity across regions in the number of infected and susceptible individuals, and (b) heterogeneity in infection rates across regions and over time. 

The first confounding factor reads as follows: if some regions start their epidemic outbreak before others, than one might expect -- all things else being equal -- that they will see a larger increase in confirmed cases in subsequent periods. For example, if mitigation policies are implemented by regions with a large number of early infected cases (suggesting a more advanced stage of the epidemic), then much of the estimates could be driven by mechanical increases in the number of infections, which are unrelated to the policy's effectiveness. Similarly, as the epidemic unfolds and the size of the susceptible population decreases, these regions will also be more likely to experience a decrease in infected cases before others.

The second confounding factor has been the primary focus of applied work. It can take multiple forms and generally refers to the endogeneity of the infection rate. For instance, many time-invariant determinants could explain different infection rates across regions and ultimately different epidemic trajectories (e.g., population density, age structure, cultural norms, political preferences, etc.). In addition, infection rates may vary over time if individuals react to information (e.g., daily death counts, research contributions, rumors, etc.) and policies (e.g., lockdowns, mask mandates). Identifying the relevant variables to account for these channels is challenging, and unobserved heterogeneity in infection rates presents a serious threat to causal inference. 

Both types of confounding factors are not merely theoretical. Relying on Covid-19 confirmed infections and mandatory lockdowns in the United States between February and June 2020, I present three pieces of supporting evidence in this direction. First, there was substantial heterogeneity across states, both in the size of their infected populations and in their baseline infection rates. Second, the timing of mandatory lockdowns is correlated to both variables, as states with tougher epidemic outbreaks implemented lockdowns earlier. Third, even states which did not implement lockdowns experienced a large decrease in infections by June 2020, suggesting other time-varying confounders played a role in slowing down the epidemic.

To isolate the effect of mitigation policies from the effects of such confounding factors, multiple studies have relied on difference-in-differences (DID) designs. DID compares the outcomes over time of regions that implement a mitigation policy to the outcomes of regions who do not. Under the assumption that treated and untreated regions would have followed parallel trends in outcomes in the absence of treatment, it is well-known that the DID estimator recovers unbiased treatment effect estimates. Unfortunately, under the SIRD model, I find that the assumption of parallel trends generally does not hold (precisely because of the two types of confounding factors discussed previously). Furthermore, apparent common trends in the pre-treatment period are as likely as they are misleading. Regions with heterogeneous infection rates usually display similar trajectories in the early stages of an epidemic, but their outcomes eventually diverge as the epidemic unfolds. These observations make DID designs less attractive in this context than in many other settings in economics. 

In practice, researchers rely on FE linear regression models for estimation. This raises two additional concerns. First and foremost, the assumption of linear additive effects is inconsistent with the SIRD model. The intuition for this is that region-specific heterogeneity non-linearly affects the entire shape of the epidemic curve. As a result, even time-invariant heterogeneity in infection rates (the unobservable) results in time-varying heterogeneity in confirmed cases (the dependent variable). This runs counter to the assumption that region-specific (but time-invariant) and time-specific (but region-invariant) heterogeneity are additively separable. This issue also extends to conditioning on a set of observed time-varying covariates which affect confirmed cases through their effect on the infection rate. A second concern -- though perhaps more obvious and less troublesome for applied work -- is that the SIRD model predicts heterogeneous treatment effects over time and across regions. In turn, failing to account for dynamic treatment effects may result in heavily biased estimates.

Overall, FE models tend to be misspecified. But how big of an issue is this in practice? SIR-type models are often solved for numerically, which complicates the formal characterization of the resulting bias in estimates. To get a sense of its magnitude, I simulate datasets from a SIRD model. In the simulations, I allow for a direct decreasing effect of mitigation policies on the infection rate. I also allow for unobserved heterogeneity in the infection rate across regions. All other endogeneity channels are shut down: there are no spillovers, the timing of mitigation policies is random and testing capacity is such that all infected individuals are observed. In this hypothetical `best of worlds', I study two estimators: the classical multi-period FE linear regression assuming constant treatment effects (i.e., a `classical DID'), and a more flexible FE model that allows for dynamic treatment effects (i.e., an `event study'). When the policy has no effect on the infection rate, both specifications can result in statistically significant estimates. When the policy decreases the infection rate, estimates based on the assumption of constant treatment effects are particularly erroneous. Event study estimates perform relatively better, especially when it comes to estimating the policy's effect `on impact' (i.e., the first period after its implementation). Nonetheless, their resulting counterfactuals are off by a large margin as they fail to capture unobserved heterogeneity in infection rates across regions.

The rest of this paper is organized as follows. Section \ref{sec:lit} briefly reviews the literature. Section \ref{sec:PolicyEvaluation} discusses policy evaluation in the context of a pandemic, with a special focus on observables and confounding factors. Section \ref{sec:DiscussionDID} explicitly links the original difference-in-differences estimator and the SIRD model of disease transmission. Section \ref{sec:DiscussionFE} discusses additional issues raised by the use of FE models for estimation. Section \ref{sec:monte_carlos} presents the results of the simulations. Finally, Section \ref{sec:Conclusion} concludes and discusses avenues for future research.

\section{Related Literature} \label{sec:lit}

Recently, multiple studies have relied on reduced form estimations to assess the causal impact of mitigation policies for Covid-19 (e.g. face mask mandates, lockdowns, quarantines) on health outcomes (e.g., confirmed infected cases, deaths). They exploit the staggered implementation of mitigation policies across regions to estimate treatment effects and conduct counterfactual analysis. This is a familiar setup for empirical work in economics, which intuitively suggests the use of difference-in-differences methodology \citep{goodman2020using}. In practice, most studies work with FE linear regression models \citep{hsiang2020effect, dave2021shelter,dave2020jue,fowler2020effect,villas2020we,courtemanche2020strong,lyu2020community,karaivanov2020face,juranek2020effect,persico2021effects}. Alternatively, a smaller set of papers relies on synthetic control methods \citep{dave2021contagion,friedson2020did,Mitze32293,cho2020quantifying}. There is an apparent growing consensus that mitigation policies significantly reduce the spread of the disease. Though estimations vary drastically across settings and methodologies, the effects uncovered are generally sizable. For instance, \cite{hsiang2020effect} estimate mitigation policies implemented between February and April 2020 prevented (or delayed) approximately 61 million confirmed cases and 495 million infections in a sample of six countries.\footnote{Though much of the focus has been on mitigation policies, it is worth noting several contributions have also studied the impact of information (or misinformation) on health outcomes \citep{ash2020effect,bursztyn2020misinformation}. They rely on a different empirical strategy for causal identification -- namely the instrumental variables approach -- but ultimately also specify a FE linear regression model.}

To bridge the divide between reduced-forms and structural approaches, some contributions have combined structural econometrics with variants of SIR models to evaluate the efficiency of mitigation policies. On the one hand, \cite{chernozhukov2020causal} build on Directed Acyclical Graphs (DAGs) to explicitly model how information, policies and behavioral responses dynamically determine the spread of the disease. The resulting empirical specification is motivated by the SIRD model, but ultimately makes different parametric assumptions. The authors highlight the importance of information and behavioral responses, but nonetheless conclude that mitigation policies effectively reduced the spread of the disease in the United States. On the other hand, \cite{allcott2020economic} explicitly combine a reduced-form estimation with a simplified SIRD model. They find modest mitigation policy effects and argue that most social distancing is driven by voluntary responses in the United States. The authors note that event studies (combined with region-specific linear time-trends) may give bizarre results in the context of an epidemic, though they do not provide a formal explanation. This paper builds on their intuition and provides simple theoretical explanations for this puzzling observation.

In a paper made public after this work, \cite{callaway2021policy} also caution against the use of DID designs for policy evaluation during a pandemic.\footnote{The first working paper version of this work was published in January 2021 (see \url{https://new.crest.science/wp-content/uploads/2021/01/2020-32.pdf}). \cite{callaway2021policy}'s working paper was published online on May 2021.} The rationale for their result is heterogeneity across regions in the timing of the epidemic outbreak (i.e., the first confounding factor I raise in the introduction). They propose an unconfoundedness approach as an alternative to DID. However -- as the authors note -- if infection rates vary both across regions and time (i.e., the second confounding factor I raise in the introduction), then unconfoundedness will not suffice to recover unbiased treatment effect estimates. Finally, though most of empirical work has been conducted on log-transforms of confirmed cases, the authors consider cumulative infected individuals as a dependent variable. In this paper, to ensure the fairness of the critique, I remain closer to the specifications of previous empirical contributions.

\section{Policy Evaluation During a Pandemic} \label{sec:PolicyEvaluation}

In this section, I introduce a simple model of epidemic dynamics and discuss the main threats to the identification of treatment effects.

\subsection{A Canonical Model of Disease Transmission}

I consider a discrete-time Susceptible-Infected-Recovered-Dead (SIRD) model (\cite{kermack1927contribution}; see \cite{allen2008introduction} for stochastic versions). Though the SIRD model is not necessarily the ground truth of how an epidemic unfolds, it is the canonical model in epidemiology to understand the hump-shaped curve of infections. As such, I see it as a useful baseline to think of epidemic dynamics and confounding factors.

\paragraph{Epidemic Dynamics.} Denote $S_{i,t}$, $I_{i,t}$, $R_{i,t}$, and $D_{i,t}$ the number of susceptible, infected, recovered, and deceased individuals at time t for region $i$. The population may be written:
\begin{equation*}
    N_i = S_{i,t} + I_{i,t} + R_{i,t} + D_{i,t}
\end{equation*}

For all periods $t$, dynamics of the epidemic are modeled as follows: 
\begin{align}
    I_{i,t+1} & = I_{i,t} + \beta_{i,t} I_{i,t} \frac{S_{i,t}}{N_i} - \gamma I_{i,t} \label{eq:I_t}\\
    S_{i,t+1} & = S_{i,t} - \beta_{i,t} I_{i,t} \frac{S_{i,t}}{N_i} \label{eq:S_t} \\
    R_{i,t+1} & = R_{i,t}  + (1-\mu) \gamma I_{i,t} \label{eq:R_t} \\
    D_{i,t+1} & = D_{i,t} + \mu \gamma I_{i,t} \label{eq:D_t}
\end{align}

$\beta_{i,t}$ is the infection rate, $\gamma$ is the inverse of the recovery time for infected individuals and $\mu$ is the mortality rate of the disease. 

In the SIRD system, the outcomes at each period are entirely determined by the preceding period's outcomes. Three things happen at each period $t$. First, infected individuals transmit the disease to a fraction of individuals who are still susceptible of catching the disease at a rate $\beta_{i,t}$. Second, a fraction $\gamma (1-\mu)$ of infected individuals recover from the disease and become immune. Third, a fraction $\gamma \mu$ of infected individuals dies from the disease. Equations \ref{eq:I_t}, \ref{eq:S_t}, \ref{eq:R_t} and \ref{eq:D_t} respectively determine the size of the infected, susceptible, recovered and deceased populations.

Consider now Equation \ref{eq:I_t} in greater detail. In the early-stages of the epidemic, the susceptible population $S_{i,t}$ is close to the size of the entire population $N_i$. The number of infections is thus relatively unimpeded and grows exponentially, as captured by the well-known basic reproduction number:
\begin{equation*}
    \mathcal{R}_{0,i,t} = \frac{\beta_{i,t}}{\gamma}
\end{equation*}

However, exponential growth is short-lived. As the epidemic unfolds, the size of the susceptible population decreases as a larger fraction of individuals becomes immune to the disease. In turn, this slows down the growth of infections up to a point where the disease ceases to spread. This results in the infamous hump-shaped curve of infections and deceased individuals.

In this context, the infection rate $\beta_{i,t}$ is of particular interest for two reasons. First, in the absence of medical breakthroughs, it drives variations in the basic reproduction number. Second, it is likely an endogenous variable in the model. Note that even a one-off shift in the infection rate will non-linearly affect the shape of the epidemic curve for all subsequent periods.\footnote{To see this, consider Equations \ref{eq:I_t} and \ref{eq:S_t}: for any given period $t$, the infection rate $\beta_{i,t}$ impacts the size of the infected and susceptible populations $I_{t+1}, S_{t+1}$, which then impact the size of the infected populations $I_{t+2}, S_{t+2}$, which then recursively impact all future infected and susceptible counts in the system of differential equations.}

\subsection{Treatment Effects and Dependent Variables}

\paragraph{Treatment Effects Notation} I now introduce treatment effects notation to rigorously discuss causal identification \citep{rubin1974estimating}. Consider a group of researchers observes multiple regions, indexed by $i \in \{1,...,K\}$, over multiple time periods, indexed by $t \in \{0,...,T\}$. They wish to inform a policy maker on the impact of a dichotomous mitigation policy $D_{i,t}$ (the `treatment') on health outcomes $Y_{i,t}(D_{i,t})$. The policy's causal effect may be thought of as $Y_{i,t}(1) - Y_{i,t}(0)$. In most applications, the main parameter of interest is therefore the Average Treatment Effect on the Treated (ATT):
\begin{equation} \label{eq:att}
    ATT_t = \mathbb{E}[Y_t(1) - Y_t(0) \ | \ D = 1]
\end{equation}

The $ATT_t$ measures the average difference between health outcomes under a mitigation policy and health outcomes in the absence of a mitigation policy among regions which implemented a mitigation policy. The fundamental problem of causal inference is that $Y_{i,t}(1)$ and $Y_{i,t}(0)$ are not observed simultaneously for the same region. For a treated region $i$, the researchers' main concern is therefore to find a counterfactual for $Y_{i,t}(0)$ - which I denote $\widehat{Y_{i,t}}(0)$.

\paragraph{Dependent Variables.} Unfortunately, though the infection rate $\beta_{i,t}$ is the primary endogenous variable in the model -- and the core parameter to capture epidemic dynamics -- it is generally unobserved. Researchers typically observe the number of confirmed infected cases $C_{i,t}$ as the epidemic unfolds. It is quite natural to use a transform of $C_{i,t}$ as a dependent variable, which I denote $Y_{i,t}$. In previous studies, the three preferred dependent variables are: 
\begin{align} 
    Y_{i,t} = \sum^t_{i = 1} \log(C_{i,t}) \label{eq:cumsum_c} \\
    Y_{i,t} = \log(C_{i,t})  \label{eq:log_c} \\  
    Y_{i,t} = \Delta_{(t,t-1)} \log(C_{i,t}) \label{eq:delta_log_c}
\end{align}
Equation \ref{eq:cumsum_c} is the cumulative number of confirmed cases in period $t$ Equation \ref{eq:log_c} is an approximation of the percentage change in confirmed cases $C_{i,t}$, and Equation \ref{eq:delta_log_c} is an approximation of the change in the growth rate from period $t-1$ to period $t$. Most of my remarks apply to Equations \ref{eq:cumsum_c}, \ref{eq:log_c} and \ref{eq:delta_log_c}. When necessary, I clarify which dependent variable is being considered. 

In the rest of this paper, I abstract from issues related to testing capacity, and assume the infected population is perfectly monitored by the authorities.\footnote{Obviously, testing capacity has varied across regions and over time and represents yet another challenge for causal identification, but is not the focus of this paper.} At any point in time, confirmed cases are thus equal to the flow of new infected individuals:\footnote{Note that $C_{i,t}$ is often replaced by the number of deceased patients, $D_{i,t}$. The remarks from this paper also apply to dependent variables using a transform of deceased individuals.}
\begin{equation} \label{eq:C_{i,t}}
    C_{i,t} = \frac{\beta_{i,t} I_{i,t} S_{i,t}}{N_i}     
\end{equation}

\paragraph{Main Confounders} Based on Equation \ref{eq:C_{i,t}}, it becomes clear that there are two main types of confounding factors which represent a threat for causal inference: (a) factors related to the size of the infected and susceptible populations $I_{i,t}$ and $S_{i,t}$, and (b) factors related to heterogeneity in the infection rate $\beta_{i,t}$.

The first type of confounding concerns the timing of the epidemic for different regions and the initial number of infected individuals, $I_{i,0}$. From Equation \ref{eq:I_t}, we know that the number of infected individuals in region and period $t+1$ increases by a factor of $\frac{1}{N_i}\beta_{i,t} I_{i,t} S_{i,t}$. In the early stages of the pandemic, the susceptible population $S_{i,t}$ is close to $N_i$, thus regions with larger amounts of infected individuals should mechanically expect larger increases in the number of infected individuals relative to other regions, irrespective of whether mitigation policies are implemented. To make meaningful comparisons between regions, one would thus need to compare regions at similar stages of their epidemic trajectory.

The second type of confounding concerns the endogenous nature of the infection rate $\beta_{i,t}$. The mitigation policy's primary objective is to decrease the infection rate, but little is known on its other potential determinants. It is useful to think of $\beta_{i,t}$ as the product of an unknown function $\mathcal{G}$, in which $X_{i,t}$ is a vector of variables which could influence the population's behavior (e.g. the number of recorded deaths, public medical information, disinformation on social media, government announcements), and of a mitigation policy $D_{i,t}$:
\begin{equation} \label{eq:beta_t}
    \beta_{i,t} = \mathcal{G}(X_{i,t}, D_{i,t})
\end{equation}

\begin{figure}[h!]
    \begin{center}
	\includegraphics[width = 0.7\textwidth]{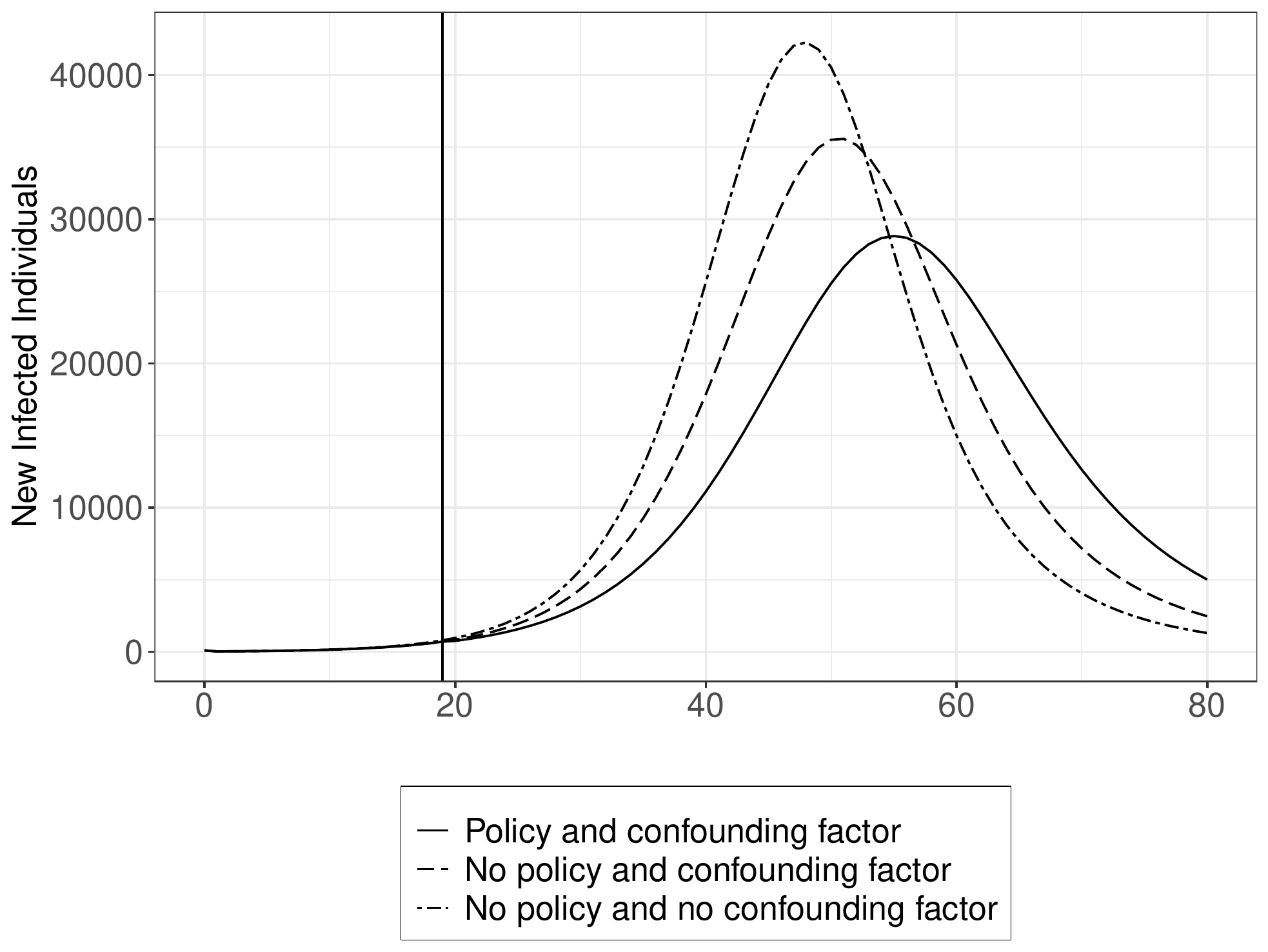}
	\caption{Treatment Effects and Confounding Factors}
	\label{fig:endogenous_beta}	
	\end{center}
	\footnotesize
	\renewcommand{\baselineskip}{11pt}
	\textbf{Note:} Figure 1 presents the evolution of an epidemic under three different scenarios: (a) a policy is implemented and there is an additional confouding factor; (b) no policy is implemented and there is an additional confounding factor; (c) no policy is implemented  and there is no confouding factor. The confounding factor increases social distancing over time with or without a mitigation policy. If an analyst assumes no confounding factor for counterfactual analysis, then the third scenario may be thought of as the estimated counterfactual $\widehat{\text{Y}}(0)$ for the first scenario $\text{Y}(1)$, and the second scenario may be thought of as the true counterfactual $\text{Y}(0)$. In this example, the analyst would largely overestimate the effect of the mitigation policy.
\end{figure}

Though researchers observe whether the policy was implemented, they do not know what would have been the infection rate without a policy implementation (as $\mathcal{G}$ is unknown). In epidemiology, researchers sometimes assume away confounding factors to conduct causal inference. For example, \cite{flaxman2020estimating} write: `\textit{Our methods assume that changes in the reproductive number – a measure of transmission - are an immediate response to these interventions being implemented rather than broader gradual changes in behaviour.}' This is a strong assumption and social scientists would expect, for instance, behavioral responses unrelated to the mitigation policy. In turn, researchers are likely to attribute part of the effects of changes in $X_{i,t}$ over time to the policy's success or failure, as portrayed in Figure \ref{fig:endogenous_beta}.

\paragraph{Supporting Evidence from the U.S.} These two types of confounding factors are not merely theoretical. In the United States, the analysis of Covid-19 confirmed infections in March 2020 (i.e., in the very early stages of the pandemic) suggests states differed in the size of their infected populations as well as in their baseline infection rates. Furthermore, the timing of treatment is correlated to both variables, as states with tougher epidemic outbreaks implemented lockdown policies earlier. On average, early-adopters of mandatory lockdowns experienced 50\% higher growth rates in infections, and had three times more cumulative confirmed infections than never-adopters. Interestingly, even states that did not implement a mandatory lockdown put an end to the first wave by June 2020. This suggests that other time-varying factors contributed to slowing down the spread of the disease (e.g., other policies, behavioral responses of the population). A more detailed analysis of U.S. data is provided in the Online Supplement.\footnote{Data on Covid-19 confirmed infections is noisy, in particular in the very early-stages of the epidemic. Nonetheless, I understand these stylized facts as suggestive evidence the concerns expressed in this paper are not merely theoretical.}

\section{Link with Difference-in-Differences Methodology}\label{sec:DiscussionDID}

In an attempt to isolate the effects of mitigation policies from the effects of confounding factors, many papers in the economics literature have exploited the staggered implementation of policies across regions and over time using difference-in-differences (DID) designs. The rationale behind DID is to compare the outcomes of treated and untreated regions, under the assumption that, albeit for the mitigation policy $D$, both groups would have experienced the same average trends in outcomes. 

\paragraph{Formal Definition.} For simplicity, I consider the case where a fraction of regions are simultaneously treated at time $t^* \in \{0,...,T\}$.\footnote{It is straightforward to extend the results to staggered implementation designs.} 

Formally, for $t \geq t^*$, the DID estimator is given by:
\begin{equation}
    DID_t = \mathbb{E}\big[ Y_t - Y_{t^*} \ | \ D = 1 \big] - \mathbb{E}\big[ Y_t - Y_{t^*} \ | \ D = 0 \big] 
\end{equation}

The parallel trends assumption reads as follows:

\begin{assumption} \label{a:paralleltrends}

    \textbf{(Parallel Trends)} ~ 
    
    \[ \forall t \in \{1,...,T\}:  \mathbb{E}\big[Y_t(0) - Y_{t^*}(0) \ | \ D = 0\big] = \mathbb{E}\big[Y_t(0) - Y_{t^*}(0) \ | \ D = 1\big] \]

\end{assumption}

Under Assumption \ref{a:paralleltrends}, it is well-known that the DID estimator provides unbiased treatment effect estimates:
\begin{equation}
    ATT_t = DID_t
\end{equation}

\paragraph{How likely is it that parallel trends hold under the SIRD model?} The short answer is that it is unlikely for any of the dependent variables used in the literature (recall Equations \ref{eq:cumsum_c}, \ref{eq:log_c} and \ref{eq:delta_log_c}).  

For example, if regions have different time-invariant baseline infection rates (e.g, population density could increase the likelihood of an infectious contact between individuals), then they will not follow parallel trends. If they have the same baseline infection rates, but different numbers of initially infected individuals relative to their respective populations (e.g, a group of infected individuals enters the country by airplane and infects the passengers of the flight in one region, compared to another region where one infected individual visits a friend and contaminates her), then they will not follow parallel trends. Similarly, if all regions have exactly the same underlying parameters, but differ in the timing of the first infected individual (e.g., the first infected hub starts in region 1 and the virus takes ten days to spread to region 2), then they will follow parallel trends. 

\begin{figure}[h!]
	\begin{center}
	\includegraphics[width = \textwidth]{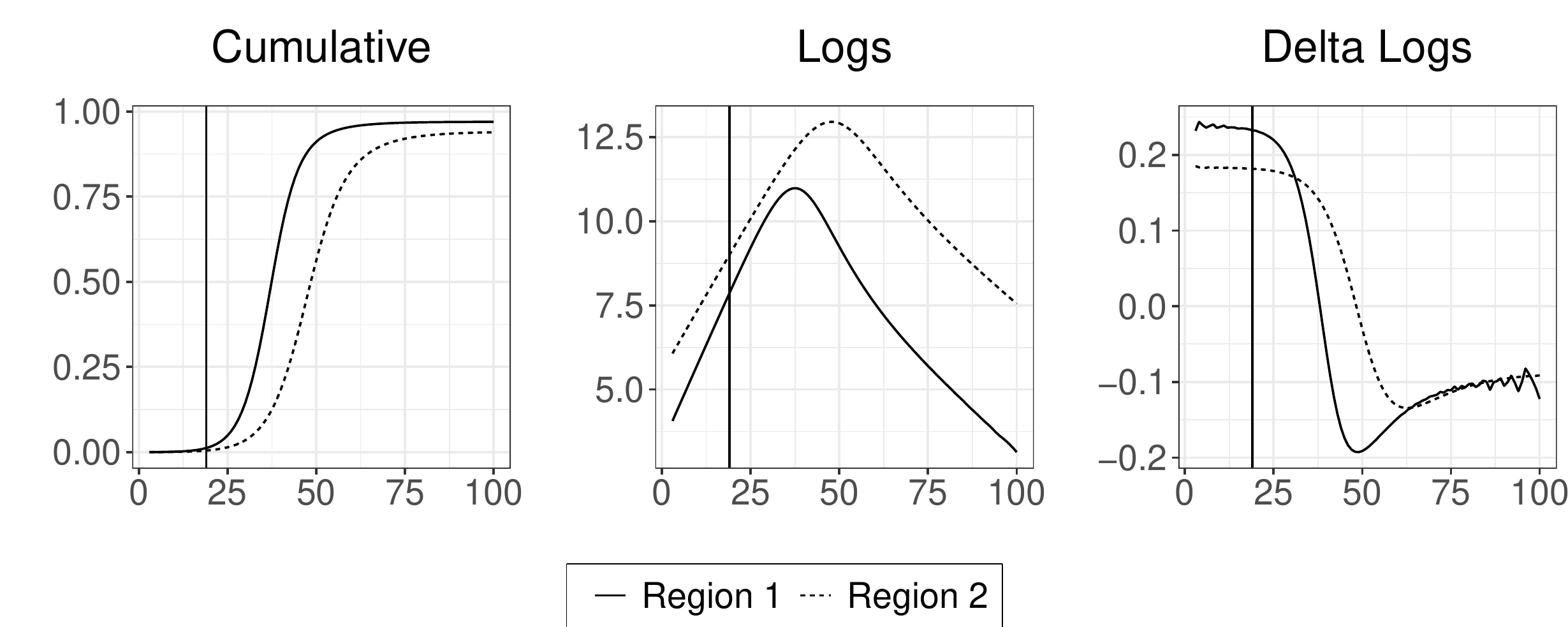}
	\caption{An Example of Apparent Common Trends}
	\label{fig:misleading_common_trends}
	\end{center}
	\footnotesize
	\renewcommand{\baselineskip}{11pt}
	\textbf{Note:} This is a graphical example of misleading common trends in logs (left panel) and delta logs (right panel). Unit 1 (black line) has a baseline contact rate $\beta = 0.3$, whereas unit 2 (grey line) has a 20\% higher baseline contact rate than Unit 1. The treatment takes place at period 20 and has a null effect on the contact rate, yet trends of both units largely diverge in the post-treatment period.
\end{figure}

In practice, given that the assumption of parallel trends is untestable, researchers usually study trends in outcomes in the pre-treatment period. If the treated and untreated regions followed common trends, this is considered as suggestive evidence that the parallel trends assumption holds. In the context of an epidemic, however, this validation heuristic can be misleading. In the early stages of the outbreak, the number of infected individuals is typically small and the susceptible population is close to the entire population size. This tends to blur \textit{ex-ante} differences across regions. Figure \ref{fig:misleading_common_trends} provides a graphical illustration of the issue.

The issue is particularly pronounced for growth rates as a dependent variable (as defined in Equation \ref{eq:delta_log_c}). Under the SIRD model, if we assume that baseline infection rates are time-invariant, two regions with different growth rates at time $t$ will by construction see their trends diverge in subsequent periods. In fact, in the early stages of the epidemic, common trends that do not overlap are a strong indication that the underlying infection rates differ between regions and that they will not experience similar epidemic trajectories -- irrespective of whether a mitigation policy is implemented.

\paragraph{Characterization of the Bias.} If common trends do not hold, then the bias is given by the following analytical formula:
\begin{align}
    DID_t - ATT_t & = \mathbb{E}\big[Y_t(0) - Y_{t^*}(0) \ | \ D = 1\big] - \mathbb{E}\big[Y_t(0) - Y_{t^*}(0) \ | \ D = 0\big]
\end{align}    
    
In simpler terms, the bias in estimates is equal to the average difference in outcomes that treated and untreated regions would have experienced in the absence of the mitigation policy. Given that SIR-type models are usually solved for numerically, the formula is rather uninformative and does not allow us to determine the sign and magnitude of the bias. 

Nonetheless, the simple case of two regions $i \in \{1,2\}$ and two periods may already provide some intuition on the role of confounders. For simplicity, let us assume that the contact rate is determined by a baseline infection rate $\beta_0$ (common to all regions), a region-specific, time-invariant component $\delta_i$ and the mitigation policy $D_{i,t}$: 
\begin{equation} \label{eq:simple_beta_t}
    \beta_{i,t} = \beta_{0} \cdot \text{exp}\Big(\delta_i + \tau D_{i,t} \Big) 
\end{equation}

In this case, assuming region 1 is treated at time $t$, taking the cumulative number of confirmed cases as a dependent variable, the bias may be written:\footnote{Similar expressions may be derived for other common transforms of confirmed cases $C_{i,t}$. The cumulative number of confirmed cases has the advantage of being the dependent variable policy-makers ultimately care about, and the dependent variable which results in the simplest analytical formula for the bias.}
\begin{equation} \label{eq:SimpleExample}
    \beta_{0}\text{exp}(\delta_1) I_{1,t} S_{1,t} -  \beta_{0}\text{exp}(\delta_0) I_{0,t} S_{0,t} 
\end{equation}

From Equation \ref{eq:SimpleExample}, the two main types of confounders appear clearly. First, larger baseline infection rates (i.e., $\delta_i$) will mechanically lead to larger amounts of infected individuals. Second, larger infected populations and lower susceptible populations (i.e., $I_{i,t}$ and $S_{i,t}$) in the pre-treatment period will also mechanically lead to larger amounts of infected individuals in subsequent periods. Both confounders will ultimately influence outcomes irrespective of whether the mitigation policy is effective.

\section{Link with Two-Way Fixed Effects Linear Models}\label{sec:DiscussionFE}

In practice, researchers rely on FE linear regression models for estimation. Contrary to a widespread belief in applied work, the classical FE model estimator is generally not equivalent to the original difference-in-differences estimator \citep{imai2020use}. Irrespective of whether the parallel trends assumption holds, the assumed data-generating process in FE models raises two additional concerns.

First and foremost, it is generally inconsistent with the SIRD model. The intuition behind this result is that heterogeneity in infection rates has a non-linear, time-dependent effect on common transforms of confirmed cases. This runs counter to the assumption of linear additive effects in FE models. As a direct consequence, FE models imperfectly capture the effect of observable confounders (e.g., population density) and cannot fully account for unobserved heterogeneity (modeled as a time-invariant, region-specific fixed effect). Second, under the SIRD model, a mitigation policy having the exact same effect on the infection rate will result in heterogeneous treatment effects across units and over time.

\subsection{Empirical Specification}

Denote $Y_{i,t}$ the outcome of interest for unit $i$ at time $t$, $D_{i,t}$ a dummy variable for the implementation of a mitigation policy. $\delta_i$ and $\delta_t$ are respectively unit-specific and time fixed effects. Finally, $\mathbf{X_{i,t}}$ is a vector of control variables. Two-way fixed effects models generally take the form:\footnote{This is the simplest specification. In event studies, the treatment dummy may also be interacted with time-windows to study dynamic treatment effects.}
\begin{equation} \label{eq:FE}
    Y_{i,t} = \delta_i + \delta_t + \tau D_{i,t} + \mathbf{X_{i,t}} \alpha  + \varepsilon_{i,t}
\end{equation}

FE models are widely used in the social sciences and in economics in particular. They provide a simple and flexible way to model a dependent variable with reliable estimation routines in many statistical softwares. One of their main advantages is that they allow researchers to both account for observed heterogeneity (through $\mathbf{X_{i,t}}$) and unobserved heterogeneity over time and across regions (through $\delta_i$, $\delta_t$). Under a FE data-generating process, the core assumption to identify treatment effects is that the error term $\varepsilon$ is orthogonal to explanatory variables in the model:

\begin{assumption} \label{a:exclusionrestriction}

    \textbf{(Zero Conditional Mean)} ~ 

    \[ \mathbb{E}\big[\varepsilon_{i,t} \ | \ \mathbf{X_{i,t}}, D_{i,t}, \delta_i, \delta_t \big] = 0 \]

\end{assumption}

In empirical work, researchers often discuss Assumption \ref{a:exclusionrestriction}. A particular attention is given to \textit{research design}. For instance, omitted variables, reverse causality, spillover effects, and other common endogeneity concerns are discussed in detail. In general, however, less attention is paid to \textit{modeling assumptions}. In particular, FE models come with the assumption that effects are separable and linearly additive.

There are many settings in economics for which this a reasonable assumption. For example, firms or individuals may differ in their unobserved idiosyncratic characteristics (captured by the unit fixed effect), but are often subject to common temporal shocks (captured by the time fixed effect). In the context of epidemics, this assumption could be realistic if researchers were to observe the basic reproduction number $\mathcal{R}_{0,i,t}$ or the infection rate $\beta_{i,t}$ over time. 

It is, however, inconsistent for common transforms of confirmed infected cases. To see this, consider as a dependent variable the log of confirmed infected cases $C_{i,t}$. Denote $\{\beta\}_{it} = \{\beta_{i1}, ..., \beta_{i,t} \}$ the entire history of contact rates until period $t$ for region $i$. Given Equations \ref{eq:C_{i,t}} and \ref{eq:beta_t}, we have:
\begin{equation} \label{eq:log_C_{i,t}}
    log(C_{i,t}) = log(\beta_{i,t}) + log(I_{i,t}(\{\beta\}_{i,t-1})) + log(S_{i,t}(\{\beta\}_{i,t-1})) - log(N_i)
\end{equation}

Even if the infection rate $\beta$ were to be time-invariant, it would still have a time-varying effect on $log(C_{i,t})$ through its non-linear impact on the respective sizes of the infected and susceptible populations, $I_{i,t}$ and $S_{i,t}$. I discuss the implications of this remark in more detail below for observed and unobserved heterogeneity across regions and over time.

\subsection{Confounders}

\begin{figure}[h!]
	\begin{center}
	\includegraphics[width = \textwidth]{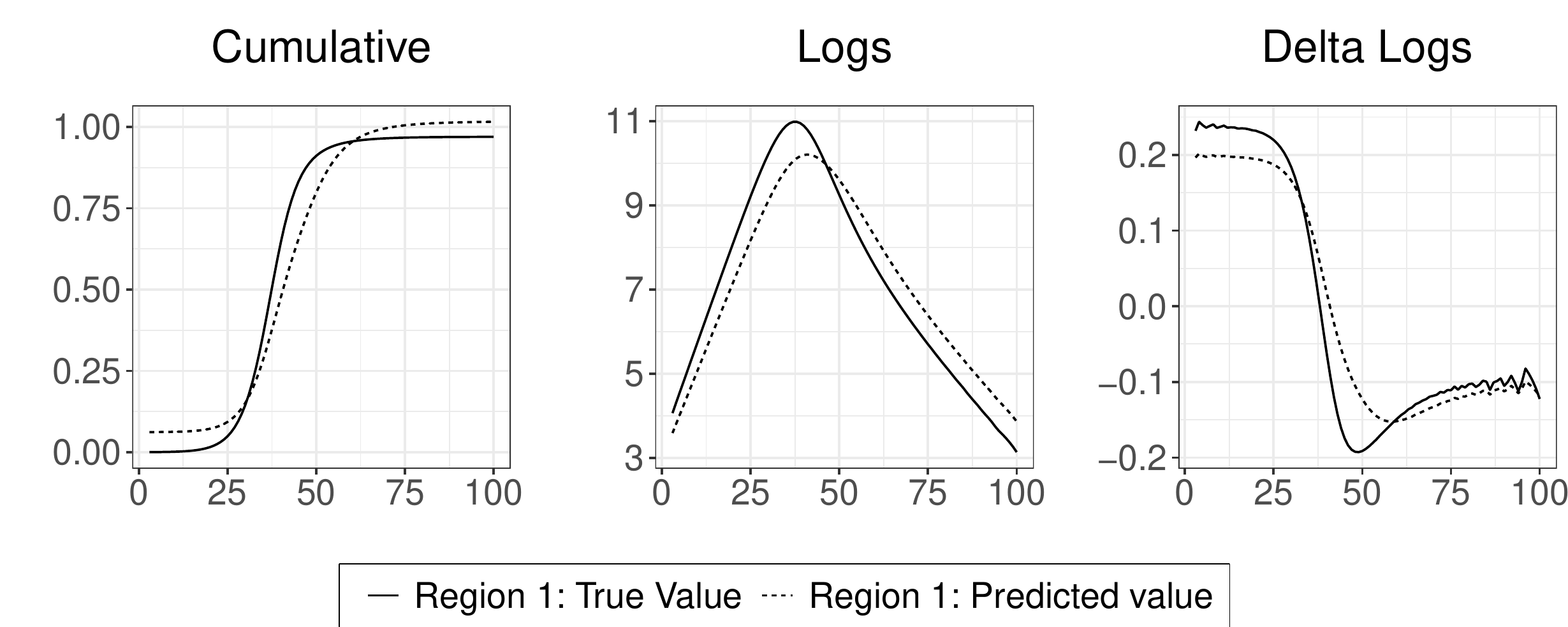}
	\caption{Fixed Effects Do Not Capture Unobserved Heterogeneity}
	\label{fig:misleading_fixed_effects}
    \end{center}
	\footnotesize
	\renewcommand{\baselineskip}{11pt}
	\textbf{Note:} I simulate the epidemic trajectories for two regions with different baseline infection rates (i.e., one region has a baseline hazard 20\% higher than the other). There is no mitigation policy. A FE linear regression model is fit on the data. I then plot the true value and predicted value for the first region based on the FE linear regression model's estimates. The model fails to capture the heterogeneity in contact rates between the two regions. Instead, time fixed effects absorb part of the resulting difference, and this results in an erroneous counterfactual for Region 1.
\end{figure}

\paragraph{Unobserved Heterogeneity.} An obvious threat to causal identification is unobserved heterogeneity across regions, which can take the form of two confounding factors. First, some regions may have started their epidemic outbreak before others and have thus a larger pool of infected individuals in the early-stages of the pandemic. Second, some regions may experience higher baseline infection rates than others (e.g, population density could increase the likelihood of an infected contact between individuals). In both cases, a larger pool of initial infected individuals or a larger baseline infection rate will produce different epidemic trajectories. Because these differences affect the flow of new infected individuals non-linearly, they cannot be captured by a region-specific, time-invariant fixed effect $\delta_i$. Figure \ref{fig:misleading_fixed_effects} presents a simple graphical illustration.

\paragraph{Observed Heterogeneity.} In practice, it is common to condition on observed potential confounders through the vector of control variables $\mathbf{X_{i,t}}$. In the context of an epidemic, potential confounders are thought to affect $C_{i,t}$ via the infection rate $\beta_{i,t}$. The concerns expressed for unobserved heterogeneity also extend to this scenario. Given Equation \ref{eq:log_C_{i,t}}, observed confounders will also have a time-varying, non-linear effect on the dependent variable through their effect on $\beta_{i,t}$. To be properly modeled, their associated coefficients $\alpha$ would need to be time-dependent, which runs counter to the FE model outlined in Equation \ref{eq:FE}.

\subsection{Heterogeneous Treatment Effects}

I now turn to the heterogeneity of treatments effects. Based on Equation \ref{eq:C_{i,t}}, a variation in the infection rate $\beta_{i,t}$ for region $i$ in period $t$ will impact infected cases in period $t+1$ by a factor of $\frac{1}{N_i} I_{i,t} S_{i,t}$. For two different regions with different sizes of infected and susceptible populations, the same mitigation policy will thus have a heterogeneous impact on new infected individuals. In addition, as mentioned previously, a change in the infection rate has a non-linear, time-dependent effect on all subsequent periods. Under the SIRD model, one thus expects mitigation policies to have heterogeneous treatment effects on infected cases across regions and over time. 

\begin{figure}[h!]
	\begin{center}
	\includegraphics[width = \textwidth]{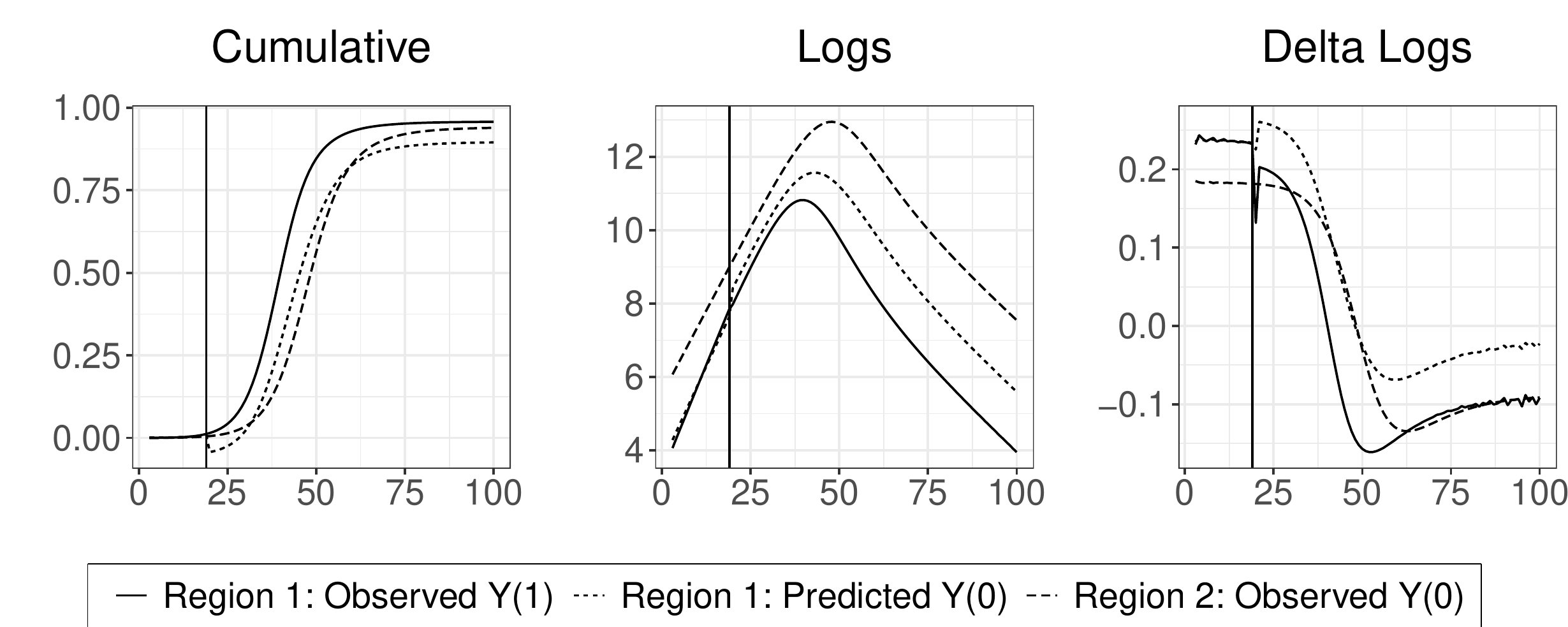}
	\caption{Assuming Constant Treatment Effects Can Bias Estimates}
	\label{fig:misleading_counterfactuals}
	\end{center}
	\footnotesize
	\renewcommand{\baselineskip}{11pt}
	\textbf{Note:} I simulate the epidemic trajectories for two identical regions. Region 1, however, is treated at time 20 and the treatment decreases the infection rate by 10\%. A classical FE linear regression model (i.e., which assumes constant treatment effects) is fit on the data. I then plot the predicted counterfactual epidemic trajectory for region 1 based on the FE linear regression model's estimates. Though common trends hold (the two regions are identical), the assumption of constant treatment effects results in a largely erroneous counterfactual.
\end{figure}

In Equation \ref{eq:FE}, the treatment effect is assumed constant over time, and failing to account for dynamic treatment effects may result in seriously erroneous counterfactuals. To illustrate this issue, a graphical example is shown in Figure \ref{fig:misleading_counterfactuals}. To mitigate this problem, it is common to specify an `event-study version' of Equation \ref{eq:FE}, in which the mitigation policy dummy $D_{i,t}$ is interacted with time fixed effects. Specifying dynamic treatment effects can solve for this problem (though not for the violation of the parallel trends assumption, nor for the inadequacy of the assumed data-generating process).\footnote{Note also that a recent body of literature has highlighted the caveats of FE models in the presence of heterogeneous treatment effects across units and over time, especially under staggered treatment implementations \citep{10.1257/aer.20181169, callaway2020difference,athey2018design,borusyak2017revisiting,goodman2018difference}. FE linear regressions estimate weighted sums of the average treatment effects (ATE) in each group and period. In such contexts, these weights can be negative and lead to biased treatment effect estimates (irrespective of whether the parallel trends assumption holds).}

\section{Simulations} \label{sec:monte_carlos}
SIR-type models are generally solved for numerically, which complicates the formal characterization of the bias. In this section, I simulate a SIRD model to get a sense of the bias and assess different specifications based on their resulting counterfactual. Obviously, one could explore the parameter space indefinitely. Nonetheless, simulations provide us with insights on how well estimators behave in well-defined theoretical contexts.\footnote{For full transparency, I provide an open-source code base for readers interested in experimenting with alternative parameter configurations. See the replication repository.} Furthermore, they are useful to understand the different dependent variables and what we should expect to observe in real-data contexts.

\subsection{Simulation Parameters and Functional Forms}

Data is simulated for 150 time periods and 100 geographical units. For simplicity, I assume that $\beta_{i,t}$ follows Equation \ref{eq:simple_beta_t}. $\tau$ is the effect of the mitigation policy on the infection rate. $\delta_i$ are geography fixed effects which could capture cultural and/or socio-demographic \textit{ex-ante} differences in the population. The baseline infection rate is to $\beta_0 = 0.12$, the recovery rate is $\gamma = 0.1$ and the death rate is $\mu = 0.01$.

I consider the case of a staggered implementation of a mitigation policy drawn at random from a uniform distribution $t_i^* \sim \mathcal{U}(1,150)$. The policy reduces the infection rate by alternatively 0 or 10\%. Regions have different population sizes, drawn from a uniform distribution $N_i \sim \mathcal{U}(10^6, 10^5)$. Regions differ in their baseline infection rates and this unobserved heterogeneity is drawn from a uniform distribution $\delta_i \sim \mathcal{U}(0,0.5)$. The unobserved heterogeneity is meant to represent the impact of time-invariant factors which increase the probability of infecting others (e.g., population density). I choose an order of magnitude in line with the case of the United States (see Online Supplement for further details). Regions also differ in their number of initially infected individuals, which is draw from a uniform distribution $I_{i,0} \sim \mathcal{U}(0.01,0.001) \times N_i$. This captures differences across regions in the timing of the epidemic, as some experience their epidemic outbreak before others.

Note that the simulated model abstracts from common endogeneity concerns expressed in the literature. Mitigation policies are implemented at random, the authorities systematically detect all new infected individuals, and there are no spillover effects. I investigate whether, in this ideal setup, the estimators perform as expected.\footnote{Explicitly correlating unobserved heterogeneity and treatment timing -- as it seems to have been the case in the United States -- would amplify the bias in the simulations.} Finally, note that the simulations are deterministic: once the parameters are drawn at random, there is no stochastic noise in the data-generating process. This implies that confidence intervals will be driven by model misspecification rather than noise.\footnote{It is straight-forward to add a stochastic component to the SIRD model through the number of infected and deceased individuals at each period. Typically, stochastic SIRD models assume that the flow if infected individuals follows a Poisson distribution $I_{i,t+1} \sim \mathcal{P}\big(\frac{\beta_{i,t} I_{i,t} S_{i,t}}{N_i}\big)$. I also run simulations with stochastic noise. This does not qualitatively change my results.}

\subsection{What should we expect for each dependent variable?} \label{DepVarSim}

Based on the simulations, each dependent variable presents different observable patterns. Upon implementation of an effective mitigation policy, logs and growth rates of infected individuals display a variation which is roughly the size of the decrease in the theoretical infection rate. However, this shift is short-lived for growth rates and lasts one period, whereas it persists over multiple periods in the case of logs. Finally, the decrease in infection rates leads to a downward change in the slope of the cumulative of new infected individuals. 

\subsection{How well do FE models perform?} \label{sims}

Though multiple dependent variables have been used in the literature (see Equations \ref{eq:log_c}, \ref{eq:delta_log_c} and \ref{eq:cumsum_c}), it remains unclear how these dependent variables relate to core epidemiological parameters such as the basic reproduction number $\mathcal{R}_{0,i,t}$. In the early-stages of an epidemic, logs and growth rates in cases are mainly driven by variations in infection rates $\beta_{i,t}$, but this ceases to be the case as the epidemic unfolds and the susceptible population decreases. As policy makers are ultimately interested in the cumulative number of infected individuals averted by mitigation policies, and since most empirical work has provided policy makers with a counterfactual along these lines, I systematically assess estimators based on their resulting counterfactual for the cumulative number of confirmed infected individuals. I compute the root mean squared error for each specification based on the number of cumulative new infected individuals. Results are summarized in Table \ref{tab:simresults}.

\paragraph{Constant Treatment Effects (i.e., `DID')} I first consider the FE model with constant treatment effects. Clearly, the treatment effect estimates are at odds with the policy's true effect on the contact rate. For example, though the policy decreases the infection by 10\%, the FE model with constant treatment effects estimates an average treatment effect close to zero, both for growth rates and the cumulative as dependent variables. When the policy is inefficient and has no effect on the infection rate, we find a positive and significant 8\% increase in infections for logs as a dependent variable. In all cases, despite the absence of noise in the simulations, the root mean squared error (RMSE) is always large, ranging from 29,000 up to 213,000 (for comparison, recall that the size of the population ranges from 100,000 to one million inhabitants per region).

\paragraph{Dynamic Treatment Effects (i.e., `event study')} I then turn to the FE model with dynamic treatment effects. Overall, event-study estimates perform better than a model assuming constant treatment effects. They notably capture the effect of the mitigation policy `on impact' relatively well, in the sense that estimates are close to the decrease in the theoretical infection rate for logs and growth rates as a dependent variable. This is true despite the presence of unobserved heterogeneity in baseline infection rates across regions (which implies a violation of the parallel trends assumption). However, the final predictions of the model result in erroneous counterfactuals which demonstrates the model's inability to successfully account for unobserved heterogeneity. Once again, despite the absence of noise in the simulations, the root mean squared error (RMSE) ranges from 34,000 to 224,000. 

\paragraph{General Take-Away} To summarize, there is a strong case against the use of FE models which assume constant treatment effects. Event-studies appear more robust, as they relax this assumption and allow for dynamic treatment effects. They notably capture relatively well the effect of the mitigation policy `on impact', even in the presence of unobserved heterogeneity in baseline infection rates across regions. Nonetheless, their resulting counterfactuals are unreliable, as their estimates are partly driven by unobserved heterogeneity across regions.\footnote{Note that I estimate large standard errors despite the absence of noise in the simulations. The standard errors reflect model misspecifications.}

\vspace{0.5cm}

\renewcommand{\arraystretch}{1.3}
\begin{table}[h!] 
\begin{center}
\begin{tabular}{c|l|l|ccc}
\hline
Policy                       & Model       & Dependent Variable & Estimate & S.E. & RMSE       \\ \hline
\multirow{6}{*}{Inefficient (0\%)} &  DID & Cumulative & 0.0117 & 0.0027 & 29949.3922 \\ 
 & DID & Log & 0.0848 & 0.0226 & 102105.3559 \\ 
 & DID & Delta Log & -0.0025 & 0.0006 & 213778.9838 \\
\cline{2-6}
 & Event Study & Cumulative & 0.0004 & 0.0075 & 34069.9416 \\ 
 & Event Study & Log & 0.0019 & 0.0748 & 101639.7259 \\ 
 & Event Study & Delta Log & 0.0003 & 0.0019 & 224595.3068 \\ 
                             \hline
\multirow{6}{*}{Efficient (-10\%)} & DID & Cumulative & -0.0077 & 0.0026 & 39123.6259 \\ 
& DID & Log & -0.1633 & 0.0219 & 102863.3005 \\ 
 & DID & Delta Log & -0.0090 & 0.0006 & 198567.7457 \\ 
\cline{2-6}
 & Event Study & Cumulative & -0.0000 & 0.0072 & 34650.7758 \\ 
 & Event Study & Log & -0.0982 & 0.0703 & 94776.0034 \\ 
 & Event Study & Delta Log & -0.0999 & 0.0018 & 209971.6188 \\ \hline
\end{tabular}
\vspace{0.2cm}
\caption{Summary of Simulation Results}
\label{tab:simresults}
\end{center}
\footnotesize
\renewcommand{\baselineskip}{11pt}
\textbf{Note:} I estimate two models on all three dependent variables used in the literature: a FE linear regression model with constant treatment effects, and a dynamic FE linear regression model (an `event study'). I also consider two scenarios. In the first scenario, the mitigation policy is inefficient and does not affect the infection rate. In the second scenario, it decreases the infection rate by 10\%. For each specification, I present the point estimate, its associated standard error, and the root mean squared error. For clarity, the latter is always computed based on the number of cumulative new infectious individuals. In the case of event study estimates, the point estimate is the estimated effect of the mitigation policy `on impact'. For completeness, the event-study graphs are provided in Appendix A.
\end{table}

\section{Concluding Remarks} \label{sec:Conclusion}

The Covid-19 pandemic has led to a vast research effort to provide policy-makers with clear and reliable take-aways. Given the economic and social costs implied by several mitigation policies, rigorously assessing their effectiveness is critical (for the pressing Covid-19 health crisis, but also more generally because other pandemics are bound to occur again).

In this paper, I highlighted several caveats of FE linear regression models to assess the impact of mitigation policies. Building on the SIRD model, I showed that FE models tend to be misspecified for three reasons: (a) the violation of the parallel trends assumption, (b) the inadequacy of the assumed data-genering process, and (c) the presence of heterogeneous treatment effects (over time and across regions). As a direct consequence, FE models generally result in biased treatment effect estimates. I provided evidence of these shortcomings on simulated datasets, and motivated these concerns with U.S. data in the early stages of the Covid-19 outbreak. Overall, my results caution against the use of difference-in-differences to study the impact of mitigation policies on health outcomes.

In many ways, these findings are reminiscent of old epistemological debates between theory-driven and atheoretical approaches to data. When we correctly understand underlying mechanisms, theory can be useful and policy relevant. I conclude that a promising line of research is for economists to build causal frameworks upon the existing structural modeling literature in epidemiology.

\section*{Acknowledgements}

I would like to thank Elliott Ash, Léa Bou Sleiman, Levi Boxell, Pierre-Edouard Collignon, Xavier d'Haultfoeuille, Antoine Ferey, Lucas Girard, Michael Knaus, Martin Mugnier, Louis-Daniel Pape, Claudia Persico, Anasuya Raj,Alessandro Riboni, Pauline Rossi, Philine Widmer and Yanos Zylberberg for their thoughtful comments. This paper benefited from fruitful discussions during presentations at CREST, ETH Zürich and University of Zürich. All mistakes are my own. 

A public repository with the code and data may be found at \url{https://gitlab.com/germain.gauthier/covid-two-way-fixed-effects.git}. 

An earlier version of this work was published in the CREST working paper series: \url{https://new.crest.science/wp-content/uploads/2021/01/2020-32.pdf}.


\bibliographystyle{apalike}
\bibliography{sample}

\appendix

\clearpage

\section*{Appendix A: Additional Material for the Simulations} \label{app:sims}

\begin{figure}[h!]
	\begin{center}
	\includegraphics[width = 0.9\textwidth]{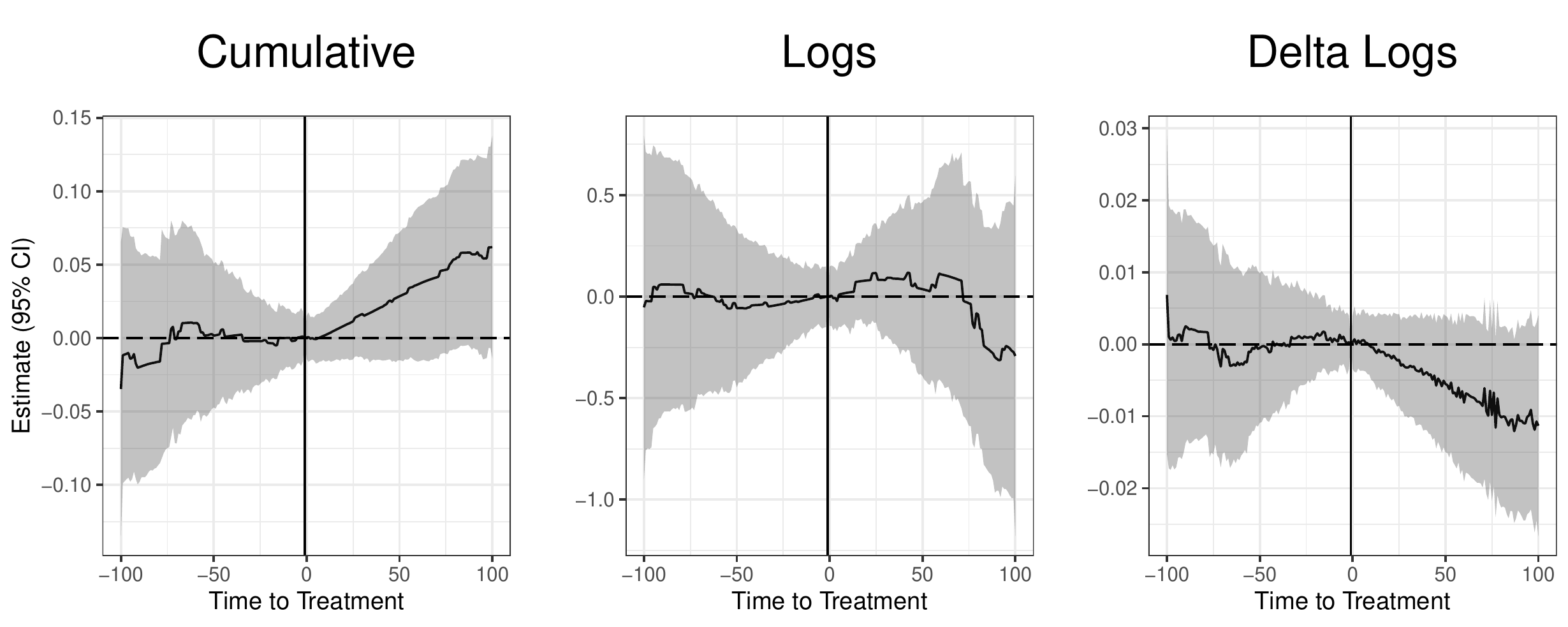}
	\caption{Event-study Estimates (Inefficient Policy)}
	\label{fig:ES0}
	\end{center}
	\footnotesize
	\renewcommand{\baselineskip}{11pt}
	\textbf{Note:} I plot treatment effect estimates resulting from an event study, taking respectively the cumulative function, a log-transform and a delta-log-transform as a dependent variable. Confidence intervals at the 95\% level are plotted in grey. The policy has no effect on the infection rate.
\end{figure}

\begin{figure}[h!]
	\begin{center}
	\includegraphics[width = 0.9\textwidth]{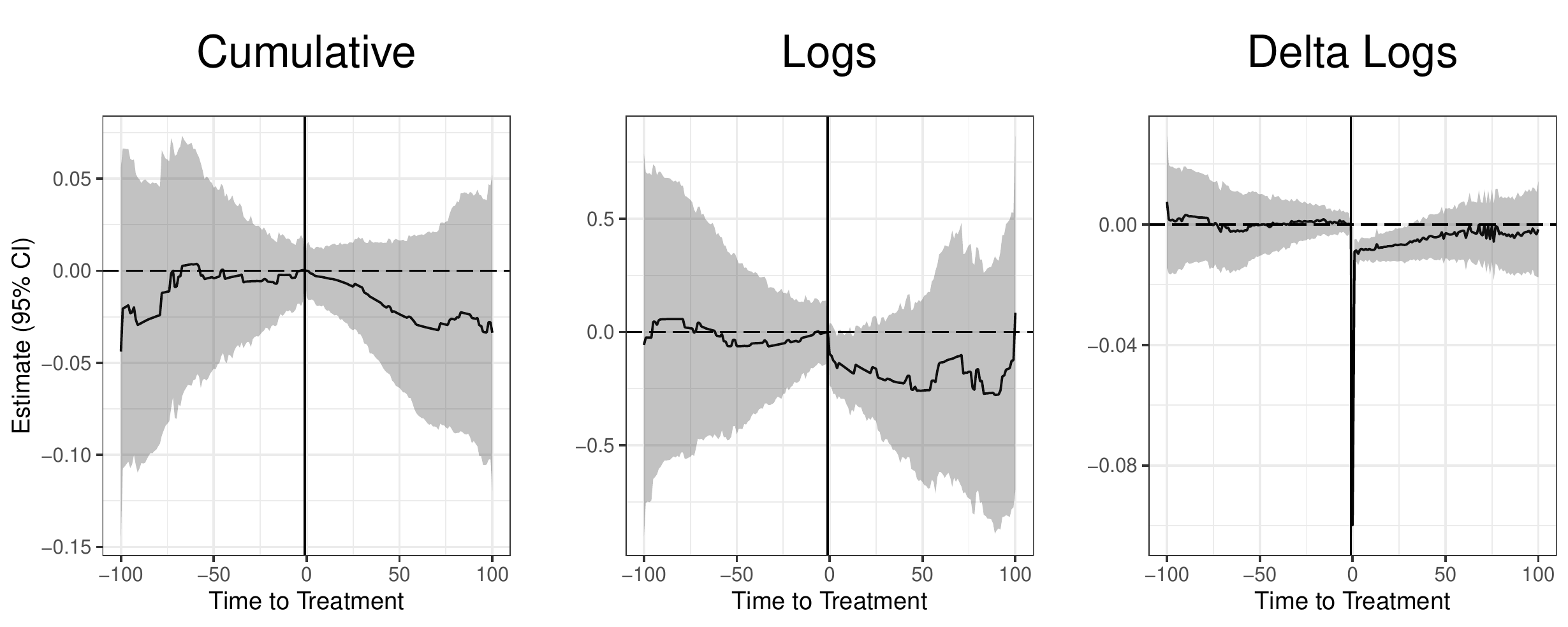}
	\caption{Event-study Estimates (Efficient Policy)}
	\label{fig:ES10}
	\end{center}
	\footnotesize
	\renewcommand{\baselineskip}{11pt}
	\textbf{Note:} I plot treatment effect estimates resulting from an event study, taking respectively the cumulative function, a log-transform and a delta-log-transform as a dependent variable. Confidence intervals at the 95\% level are plotted in grey. The policy decreases the infection rate by 10\%.
\end{figure}

\clearpage

\section*{ONLINE SUPPLEMENT \\ \vspace{0.3cm} Supporting Evidence for the United States} \label{app:us_data}

I present suggestive evidence that the issues outlined in this paper represent serious threats to causal identification in real data contexts.

\subsection{Data Sources}

I rely on two main sources of data. The first source provides the periods of time during which U.S. states -- or a subset of a state -- were under lockdown. This is measured by the ``full or partial closure of non-essential retail, ordered by local government'' as reported by Aura Vision\footnote{Data available at \url{https://auravision.ai/Covid19-lockdown-tracker/}}. The second source reports the cumulative number of infected and deceased individuals per U.S. State per day in relation to Covid-19. The data is provided by \textit{The Covid Tracking Project}.\footnote{Data available at \url{https://Covidtracking.com/data/download}} I focus on the first wave of Covid-19 in the United States. I thus limit the sample to the period starting on February 15th and ending on June 30th 2020. Figures \ref{fig:US_RAW} and \ref{fig:US_lockdowns} in the Appendix present a graphical overview of the data. 

\subsection{Stylized Facts}

I find suggestive empirical evidence that most confounding factors discussed in this paper were present in the early-stages of the U.S. Covid-19 epidemic. Table \ref{tab:UsStats} summarizes the main summary statistics.

\vspace{0.2cm}

\begin{table}[h!]
\begin{center}
\begin{tabular}{lcc} 
  \hline
Timing of lockdowns & Mean Growth Rate & Mean Cumulative \\ 
  \hline
First Week & 2.3453 & 1.2482 \\ 
Second Week & 1.8432 & 0.7998 \\ 
Third Week & 2.2141 & 0.3183 \\ 
Never & 1.5341 & 0.4258 \\ 
   \hline
\end{tabular}
\caption{Heterogeneity Across U.S. States}
\label{tab:UsStats}
\end{center}
\footnotesize
\renewcommand{\baselineskip}{11pt}
\textbf{Note:} This table presents the mean growth rate as well as the mean cumulative number of confirmed infected cases on the $22^d$ of March 2020. Growth rates are computed with the delta of the logs on a seven day basis. Cumulative counts are shown per 10,000 inhabitants. I classify states into four groups: the states which first implemented lockdowns between the $17^{th}$ and the $22^{th}$ of March (`first week'), states which implemented lockdowns in the week after (`second week'), states which implemented lockdowns two weeks after (`third week'), and states which did not issue a lockdown (`never').

\end{table}

\paragraph{Timing of the Epidemic} Out of 52 states, 7 implemented lockdowns before the $22^d$ of March (week 1), 24 in the following week (week 2), and 17 in the week after (week 3). Only 4 states did not implement mandatory lockdowns (never-takers). The mean cumulative for each group indicates that states were at different stages of the epidemic on the $22^{d}$ of March 2020. For instance, the first-movers had three times more cumulative cases per 10,000 inhabitants than the never-takers (see Table \ref{tab:UsStats}). It follows that -- irrespective of whether a lockdown is implemented -- we would expect the first-movers to experience larger counts of confirmed infected cases relative to states which implemented lockdowns later or never at all.

\paragraph{Baseline Infection Growth Rates} U.S. states are also heterogeneous in terms of their mean growth rates of confirmed infected cases, which suggests different baseline infection rates among the first-movers, the eventually treated and the never treated. For instance, the first-movers have growth rates in infections approximately 50\% higher than never-takers at mid-March (see Table \ref{tab:UsStats}). Once again -- irrespective of whether a lockdown is implemented -- we would expect treated states to experience larger counts of confirmed infected cases relative to states which implemented lockdowns later or never at all.

\paragraph{Time-Varying Confounders}

\begin{figure}[h!]
	\begin{center}
	\begin{tabular}{c}
	\includegraphics[width = \textwidth]{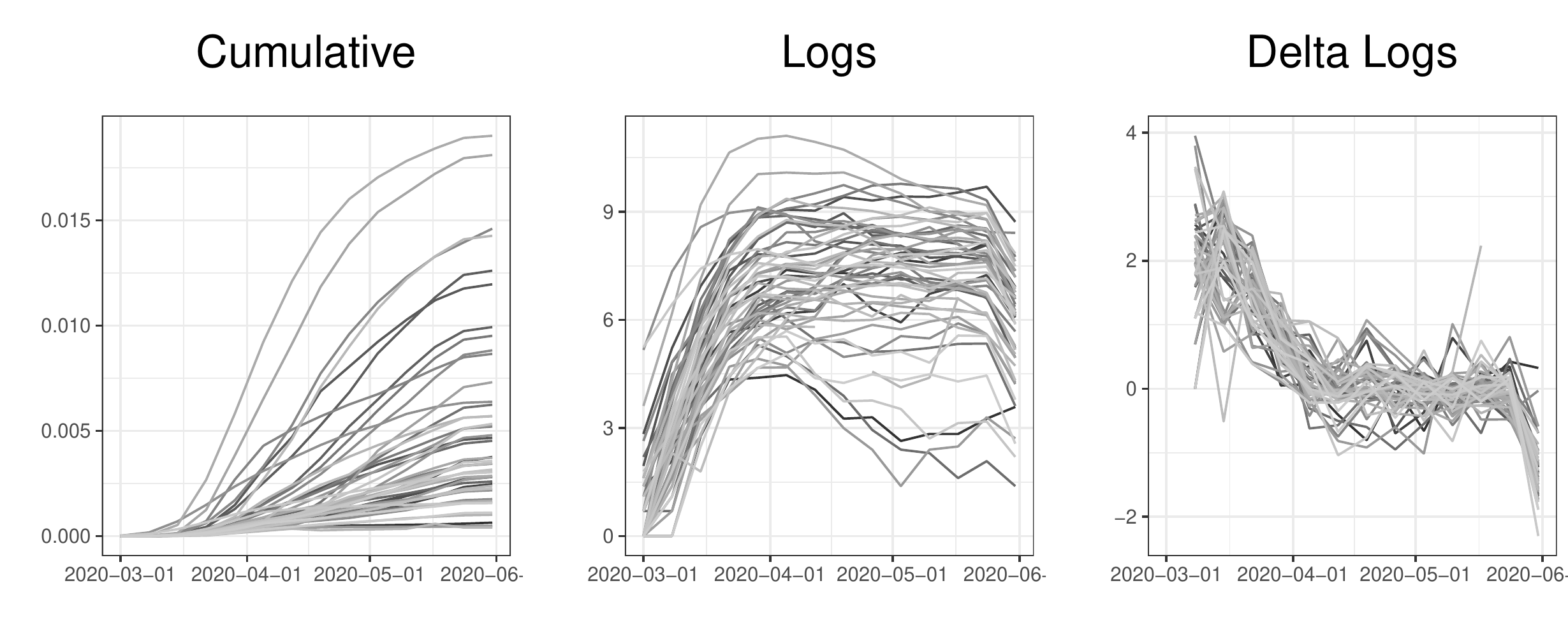}
    \end{tabular}
	\caption{U.S. Raw Data on Confirmed Cases}
	\label{fig:US_RAW}
	\end{center}
	\footnotesize
    \renewcommand{\baselineskip}{11pt}
    \textbf{Note:} I plot U.S. confirmed infectious cases taking respectively the cumulative function, a log-transform or a delta-log-transform. Each grey line represents a U.S. state.
\end{figure}

Finally, all states eventually experienced a decrease in the growth rate of confirmed infected individuals. This led to the end of the first wave (see Figure \ref{fig:US_RAW}). There are two main rationales which could explain this empirical pattern. First, effective confounding measures were taken by the authorities over the period. Second, people `learned' social distancing as the epidemic evolved.\footnote{One last explanation is that states under lockdown had positive spillover effects on untreated states. This would imply a direct violation of the Stable Unit Treatment Value Assumption (SUTVA).} Both explanations imply that time-varying confounders influenced infection rates. Those would need to be accounted for. Unfortunately, as discussed in this paper, they cannot be correctly captured by FE models. 

\begin{figure}[h!]
	\begin{center}
		\begin{tabular}{c}
		\includegraphics[width = 0.8\textwidth]{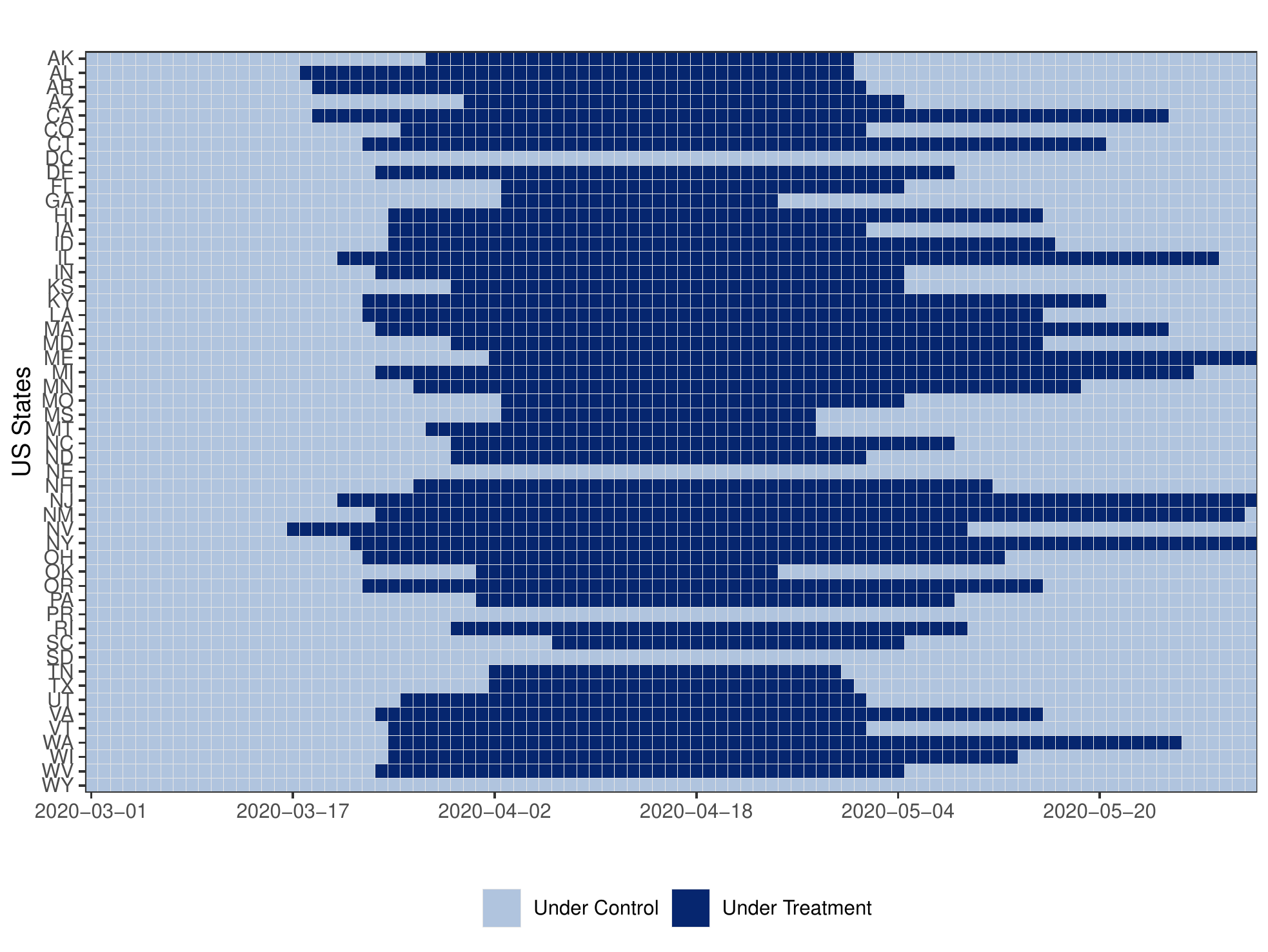}
	    \end{tabular}
		\caption{Timing of Lockdowns}
		\label{fig:US_lockdowns}
		\end{center}
		\footnotesize
        \renewcommand{\baselineskip}{11pt}
        \textbf{Note:} This is a graphical representation of the timing of lockdowns (in dark blue) in the United States for each U.S. state. 
\end{figure}

\begin{figure}[h!]
	\begin{center}
	\begin{tabular}{c}
	\includegraphics[width = \textwidth]{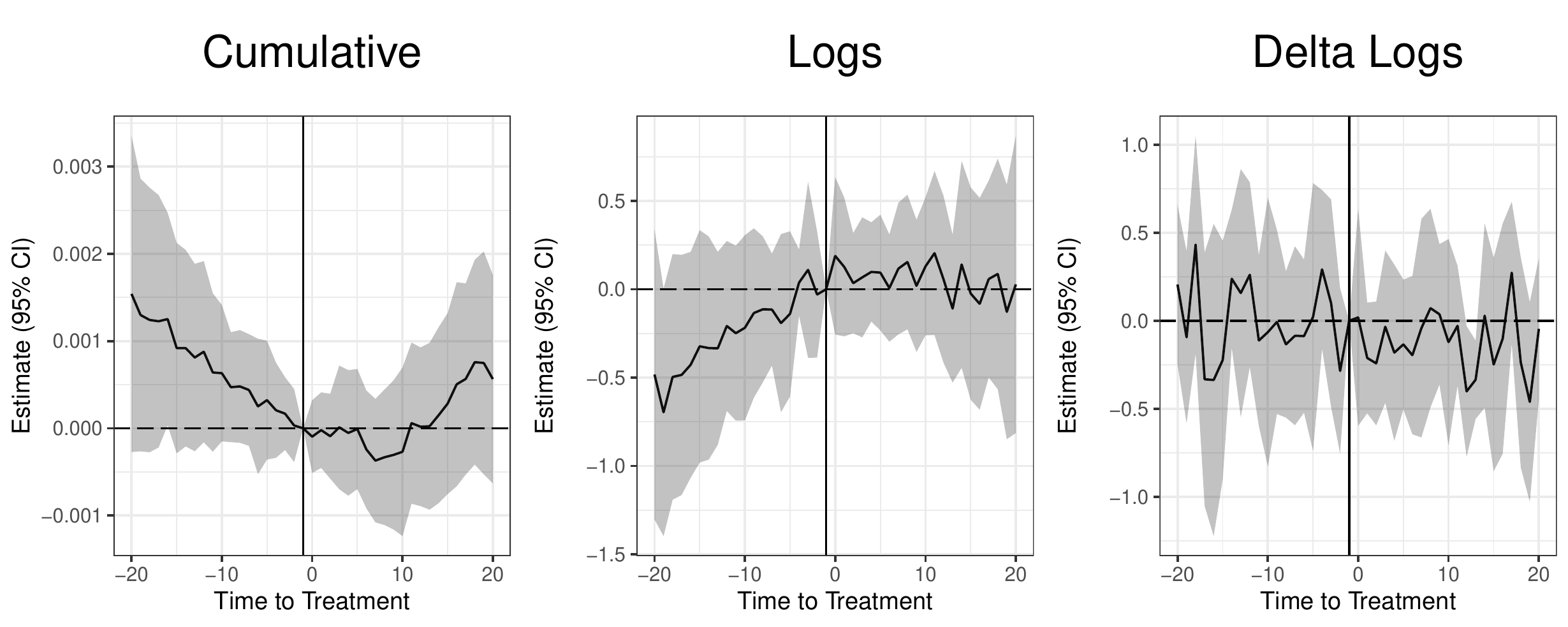}
    \end{tabular}
	\caption{Event Study Estimates on U.S. Data}
	\label{fig:ESNPT}
	\end{center}
	\footnotesize
    \renewcommand{\baselineskip}{11pt}
    \textbf{Note:}  I plot the event study coefficients based on U.S. data with a window of 40 time periods. I take respectively the cumulative function, a log-transform or a delta-log-transform as a dependent variable. Confidence intervals are at displayed in grey at the 95\% level.
\end{figure}

\end{document}